%% file: 18ASILOMAR_ULHW_arxiv.tex
\documentclass[conference,10pt,twocolumn,a4paper]{IEEEtran}

\usepackage{cite}
\usepackage[T1]{fontenc}
\usepackage{graphicx}
\usepackage{amssymb}
\usepackage{amsmath}
\usepackage{paralist}
\usepackage{setspace}
\usepackage{microtype}
\usepackage{hyperref}
\usepackage{url}
\usepackage{balance}
\usepackage[caption=false,font=footnotesize]{subfig}
\usepackage{xcolor}
\usepackage{fancyref}
\usepackage{dsfont}

\IEEEoverridecommandlockouts

\input{vmr-symbols-vecbold}
\input{standard-macros}

\input{defs}

\newcommand{\quantize}{\mathcal{Q}}

\DeclareMathOperator{\fack}{d}
\DeclareMathOperator{\diag}{diag}

\makeatletter
\def\@IEEEinterspaceratioM{0.265}
\def\@IEEEinterspaceMINratioM{0.1651}
\def\@IEEEinterspaceMAXratioM{0.38}

\def\@IEEEinterspaceratioB{0.31}
\def\@IEEEinterspaceMINratioB{0.19}
\def\@IEEEinterspaceMAXratioB{0.38}
\@IEEEtunefonts
\makeatother
\begin{document}

%

\title{Massive MU-MIMO-OFDM Uplink with \\ Hardware Impairments: Modeling and Analysis}
\author{
\IEEEauthorblockN{Sven Jacobsson$^\text{1,2}$, Ulf Gustavsson$^\text{1}$, Giuseppe Durisi$^\text{2}$, and Christoph Studer$^\text{3}$} \\ \vspace{-0.3cm}
\IEEEauthorblockA{$^\text{1}$Ericsson Research, Gothenburg, Sweden}
\IEEEauthorblockA{$^\text{2}$Chalmers University of Technology, Gothenburg, Sweden}
\IEEEauthorblockA{$^\text{3}$Cornell University, Ithaca, NY, USA}
\vspace{-0.3cm}
\thanks{The work of SJ and GD was supported in part by the Swedish Foundation for Strategic Research under grant ID14-0022, and by the Swedish Governmental Agency for Innovation Systems (VINNOVA) within the competence center ChaseOn.
The work of CS was supported in part by Xilinx Inc., and by the US National Science Foundation (NSF) under grants ECCS-1408006, CCF-1535897, CAREER CCF-1652065, CNS-1717559, and EECS-1824379.
}
}

\maketitle

\begin{abstract}

We study the impact of hardware impairments at the base station (BS) of an orthogonal frequency-division multiplexing (OFDM)-based massive multiuser (MU) multiple-input multiple-output (MIMO) uplink system. 
We leverage Bussgang's theorem to develop accurate models for the distortions caused by nonlinear low-noise amplifiers, local oscillators with phase noise, and oversampling finite-resolution analog-to-digital converters. 
By combining the individual effects of these hardware models, we obtain a composite model for the BS-side distortion caused by nonideal hardware that takes into account its inherent correlation in time, frequency, and across antennas.
We use this composite model to analyze the  impact of BS-side hardware impairments on the performance of realistic massive MU-MIMO-OFDM uplink systems.
\end{abstract}

\section{Introduction}

Massive multiuser (MU) multiple-input multiple-output (MIMO) relies on base station~(BS) architectures with hundreds of antenna elements that simultaneously serve  tens of user equipments~(UEs) in the same frequency band. 
Massive MU-MIMO promises substantial gains in terms of spectral efficiency, energy efficiency, reliability, and coverage compared to traditional small-scale MU-MIMO systems~\cite{rusek14a, larsson14a} and is hence widely believed to be a core technology in fifth generation (5G) cellular networks~\cite{boccardi14a}. 
In order to keep system costs and circuit power consumption within reasonable bounds, massive MU-MIMO has to be realized using low-cost (and, hence, nonideal) hardware components at the BS~\cite{bjornson15a, bjornson15c}.
%
%
In this paper, we  analyze an orthogonal frequency-division multiplexing (OFDM)-based massive MU-MIMO uplink system (UEs transmit to the BS) with nonideal hardware components at the BS-side.
Concretely, we analyze the joint distortion introduced  by nonlinear low-noise amplifiers (LNAs), local oscillators (LOs) with phase noise, and oversampling finite-resolution analog-to-digital converters~(ADCs). 

\subsection{Previous Work}
The impact of \emph{aggregate} hardware impairments (i.e., from multiple sources of hardware impairments) on small-scale MIMO and massive MU-MIMO systems has been investigated in, e.g.,~\cite{zhang15d, bjornson15c, studer10b}, using a statistical model that treats the distortion caused by nonideal hardware as power-dependent additive white Gaussian noise.
While such simple statistical models have been shown to yield accurate results in some scenarios when characterizing bit error rate (BER) and spectral efficiency~\cite{gustavsson14a, studer10b, bjornson18a}, they do not fully capture the impact of real-world hardware impairments on practical systems~\cite{bjornson18a, larsson18a}.
Indeed, nonideal hardware causes distortion that is, in general, correlated in time, frequency, and across the BS antenna array~\cite{larsson18a, jacobsson17a, blandino17a, moghadam12a}. 
It is therefore of practical interest to develop accurate behavioral models of nonideal hardware that capture this inherent correlation.
Behavioral models have been used previously to assess the impact  on the performance of massive MU-MIMO uplink of an isolated hardware impairment, e.g.,  nonlinearity of LNAs~\cite{mollen17b}, phase noise~\cite{wu06a, mehrpouyan12a, pitarokoilis15a, khanzadi15a}, or low-resolution ADCs~\cite{jacobsson17b, mollen16c, zhang12a, li17b}. 
A theoretical analysis with an aggregate impairment model that jointly models all these distortions is, however, missing.

\subsection{Contributions}

Starting from  behavioral models of nonideal hardware components, we derive an aggregate statistical hardware-impairment model that can be used in system-level and link-level simulations to accurately analyze the distortion caused by nonlinear LNAs, LOs with phase noise, and oversampling finite-resolution ADCs.
The proposed aggregate hardware-impairment model, which captures the inherent correlation of the induced signal distortion in time, frequency, and over the antenna array, depends only on the second-order statistics of the received signal and on the parameters of the  behavioral hardware models.
We validate the accuracy of the proposed aggregate model by comparing our analytical results with numerical simulations in a massive MU-MIMO-OFDM uplink system.
%

\begin{figure*}
\centering
\includegraphics[width = .99\textwidth]{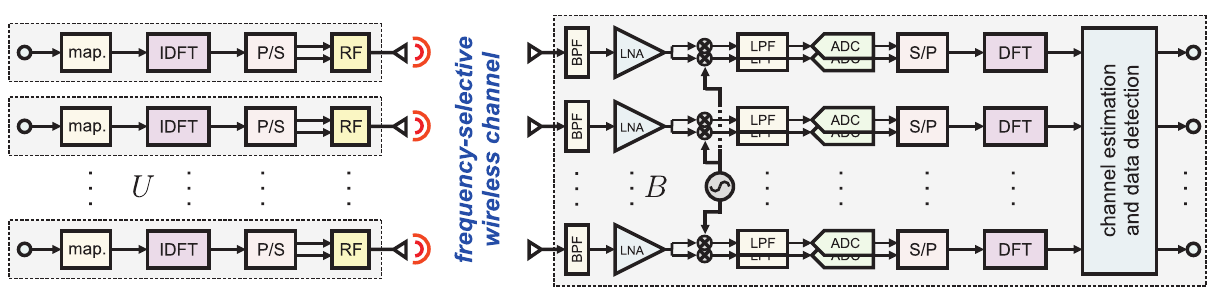}
\caption{Overview of a massive MU-MIMO-OFDM uplink system with nonideal hardware at the BS-side. Left: $U$ single-antenna UEs with ideal hardware components independently perform  symbol mapping and OFDM modulation.
Right:  The per-antenna received time-domain signal the BS is filtered using a band-pass filter (BPF), amplified using an LNA, mixed with an LO signal,  filtered with a low-pass filter (LPF), and converted into the digital domain by a pair of ADCs. After BS-side OFDM demodulation, the received signals over the $B$ BS antennas are combined using a linear filter (e.g., zero forcing) to detect the transmitted symbols.
}
\label{fig:system}
\end{figure*}

\subsection{Notation}

%
%
%
The $M \times N$ all-zeros matrix and the $M \times M$ identity matrix are denoted by $\veczero_{M \times N}$ and $\matI_M$, respectively.
%
%
The real and the imaginary parts of a complex-valued vector $\veca$ are $\Re\{\veca\}$ and $\Im\{\veca\}$, respectively. 
We use $\opnorm{\veca}_2$ to denote the $\ell_2$-norm of $\veca$. 
The main diagonal of a matrix $\matA$ is $\diag(\matA)$.  
%
%
The Hadamard product of two equally-sized matrices $\matA$ and $\matB$ is $\matA \circ \matB$. 
Furthermore, the elements of the matrix $\abs{\matA}$ are the absolute values of the elements of $\matA$.
The complex-valued circularly symmetric Gaussian distribution with covariance matrix $\matK \in \opC^{M \times M}$ is~$\jpg(\matzero_{M \times 1}, \matK)$.
The expected value of a random vector $\vecx$ is~$\Ex{}{\vecx}$. 
Finally, we let $\sinc(x) = \sin(\pi x)/(\pi x)$ for $x\neq0$ and $\sinc(x)=1$ for $x=0$.

\section{System Model} \label{sec:system}

We consider a single-cell massive MU-MIMO-OFDM uplink system as shown in~\fref{fig:system}. Here, $U$ single-antenna UEs use spatial multiplexing to communicate in the same time-frequency resource with a $B$-antenna BS.
Our model considers several nonideal hardware components at the BS, including nonlinear LNAs, phase noise, and oversampling finite-resolution~ADCs.
For simplicity, however, the analog band-pass filter (BPF) and the analog low-pass filter (LPF) at the BS-side shown in \fref{fig:system} as well as all hardware components at the UEs are assumed to be ideal.
Finally, we assume that the sampling rate $F_s$ of the digital-to-analog converters (DACs) at the UE side equals the sampling rate of the ADCs at the~BS side.

\subsection{Transmitted Signals}

We consider the practically relevant scenario in which OFDM is used to simplify equalization when operating over frequency-selective channels.
Without loss of generality, we focus on the transmission of a single OFDM symbol. Let $N$ denote the total number of subcarriers per OFDM symbol, and let $S \le N$ denote the number of occupied subcarriers. 
The subcarrier spacing is $F_\text{sub} = F_s/N$.
The oversampling rate (OSR) is defined as~$\textit{OSR} = N/S$. The case $\textit{OSR} = 1$ corresponds to symbol-rate sampling whereas the case $\textit{OSR} > 1$ corresponds to faster-than-Nyquist sampling.
We denote by $\hat{s}_u[k]$ the frequency-domain symbol transmitted from UE  $u = 1, 2, \dots, U$ on  subcarrier $k = 0,1,\dots,N-1$.
Let $\setS$ and $\setG = \{ 0, 1, \dots, N-1\} \setminus \setS$ denote the set of occupied subcarriers and guard subcarriers, respectively.
 We assume that $\hat{s}_{u}[k] = 0$ for $k \in \setG$.
With these definitions, we can write the time-domain sample $n = 0,1\dots,N-1$ of the discrete-time baseband signal transmitted from UE $u$ as follows:
\begin{IEEEeqnarray}{rCl} \label{eq:s_scalar}
s_u[n] = \frac{1}{\sqrt{N}}	\sum_{k \in \setS} \hat{s}_u[k] e^{jk\frac{2\pi}{N}n}. \IEEEeqnarraynumspace
\end{IEEEeqnarray}
To prevent interference between  OFDM symbols, a cyclic prefix of length $L-1$ is prepended to $\big\{s_u[n]\big\}$. Specifically, $s_u[n] = s_u[N+n]$ for $n = -L+1, -L+2, \dots, -1$.

Let $F_c$ denote the carrier frequency. The continuous-time passband signal transmitted from UE $u = 1, 2, \dots, U$ is then
\begin{IEEEeqnarray}{rCl} \label{eq:tx_ct_pb}
\tilde{s}_u(t) 
&=& \sqrt{2}\Re\{ s_u(t)e^{j2\pi{F_c}t}\}
\end{IEEEeqnarray}
where we use ideal sinc-based pulse shaping:
%
\begin{IEEEeqnarray}{rCl} \label{eq:tx_ct_bb}
	s_u(t) &=& \sum_{n=-L+1}^{N-1} s_u[n] \sinc\lefto(t/T_s - n\right). \IEEEeqnarraynumspace
\end{IEEEeqnarray}
Here, $T_s = 1/F_s$ is the sampling period of the DACs at the UEs, which equals the sampling period of the ADCs at the BS.

\subsection{Received Signals}

%
The received signal at each of the $B$ BS antennas is filtered using an analog BPF that, in this work, is assumed to be ideal and matched to the transmit signal.
Specifically, the filtered continuous-time passband signal received at BS antenna $b = 1, 2, \dots, B$ is  contained within the frequency interval $[F_c - F_s/2, F_c + F_s/2]$ and can be written as
\begin{IEEEeqnarray}{rCl} \label{eq:tx_ct_pb}	
	\tilde{x}_b(t) &=& \sqrt{2}\Re\lefto\{  x_b(t)e^{j2\pi{F_c}t} \right\} 
\end{IEEEeqnarray}
where the complex envelope $x_b(t)$ of $\tilde{x}_b(t)$ is given by
\begin{IEEEeqnarray}{rCl} \label{eq:tx_ct_bb}	
x_b(t) 
&=& \sum_{u=1}^U \int_{-\infty}^\infty h_{b,u}(\tau)s_{u}\lefto(t - \tau\right)\fack{\!\tau} + w_b(t). \IEEEeqnarraynumspace
\end{IEEEeqnarray}
Here, $h_{b,u}(\tau)$ is the impulse response of the linear time-invariant channel connecting the $u$th UE to the $b$th BS antenna. 
Furthermore, $w_b(t)$ is the filtered continuous-time additive white Gaussian noise (AWGN) with power spectral density (PSD) that is constant  and equal to $N_0$ in the frequency band $[F_c - F_s/2, F_c + F_s/2]$ and zero outside this~interval.

The passband signal $\tilde{x}_b(t)$ is amplified by a nonlinear LNA before being down-converted to baseband.
 The down-conversion step is performed by mixing the LNA output with a carrier signal generated by a LO.
Finally, the baseband signal is converted into the digital domain by a pair of finite resolution ADCs that perform sampling and amplitude quantization. 
Each of these operations will introduce hardware impairments to the received signal.
We next discuss  behavioral models for each source of hardware~impairments.
 
\section{Behavioral Models for Hardware Impairments} \label{sec:behave}

Behavioral models provide a simple   relationship between the input and the output of a device under test, and can be tuned to match empirical measurements.
Their key advantages compared to more sophisticated and accurate physical models are that (i) they use only a small set of parameters and (ii) they do not require deep domain knowledge of the underlying physics governing the functioning of a device~\cite{ghannouchi09a}.
Behavioral models of nonideal hardware are commonly used in the design of impairment-mitigation algorithms, such as digital predistortion and phase-noise tracking.
We next introduce behavioral models for the set of nonideal hardware components considered in this~work.

\subsection{Low-Noise Amplifier: Behavioral Model}
%
Behavioral models for nonlinear amplifiers rely typically on Volterra series~\cite{benedetto79a} or generalized polynomial models~\cite{morgan06a}, although alternatives exist~\cite{ghannouchi09a, pedro05a, tehrani10a}.
If the bandwidth of the LNAs is large compared to the bandwidth of the received signal,  the impact of the LNA on the complex envelope of the received signal can be accurately modeled as a memoryless nonlinearity that is independent of the carrier frequency~\cite{saleh81a, kaye72a}.
In what follows, we consider a simple memoryless third-order  polynomial model for which  the LNA output at BS antenna $b = 1, 2, \dots, B$ is given by 
\begin{IEEEeqnarray}{rCl}
\tilde{y}_b(t) &=& \sqrt{2} \Re\lefto\{ y_b(t) e^{j2\pi{F_c}t} \right\}
\end{IEEEeqnarray}
where
\begin{IEEEeqnarray}{rCl}
y_{b}(t) 
&=& f_\text{lna}\lefto(x_{b}(t)\right) =  \alpha_1 x_{b}(t) + \alpha_2 x_{b}(t) \lefto|x_{b}(t)\right|^2. \label{eq:f_lna}
\end{IEEEeqnarray}
Here, the coefficients $\alpha_1 \in \opC$ and $\alpha_2 \in \opC$ are the model parameters, which we assume to be known at the BS. Note that, in~\eqref{eq:f_lna}, we assumed that all LNAs at the BS are~identical.


\subsection{Phase Noise: Behavioral Model}

The output of the LNA is down-converted to baseband by mixing it with a carrier signal generated by a LO. 
The down-converted signal is then passed through an analog LPF to eliminate unwanted aliases. 
As shown in~\fref{fig:system}, we assume that all BS antennas are connected to a common LO. 
The carrier signal generated by the LO typically contains  random phase variations stemming from circuit imperfections and  noise~\cite{razavi96a, lee00a}. 
We model the filtered, down-converted signal at BS antenna $b = 1, 2, \dots, B$ as follows:
\begin{IEEEeqnarray}{rCl}
z_b(t)
&=& \text{LPF}\lefto\{ \sqrt{2} \, e^{-j2\pi F_c t + j\phi(t)} \tilde{y}_b(t) \right\} \\
&=& \text{LPF}\lefto\{ e^{j\phi(t)} y_b(t) + e^{-j4\pi F_c t + j\phi(t)}y_b^*(t)\right\} \IEEEeqnarraynumspace\\
&=& e^{j\phi(t)} y_b(t). \label{eq:f_mixer}
\end{IEEEeqnarray}
Here, $\phi(t)$ is the phase-noise process, and we have assumed that the LPF removes all aliases perfectly.

As we will discuss in~\fref{sec:composite}, we are interested in the realizations of the phase-noise process at the sampling time instances. Let $\phi[n] = \phi(nT_s)$. We model $\{ \phi[n]\}$ as the zero-mean random process defined by 
\begin{IEEEeqnarray}{rCl}
\phi[n] &=& \lambda\phi[n-1] + \varphi[n]. \label{eq:phase_noise}
\end{IEEEeqnarray}
Here,  $0 < \lambda \leq 1$ and $\varphi[n] \sim \normal(0, \sigma_\varphi^2)$. The variance of the phase innovation (the phase-noise rate) is $\sigma_\varphi^2 = 2\pi\beta T_s$. We assume that the parameters of the behavioral model, i.e., the constants $\lambda$ and $\beta$, are known to the~BS.

Some comments are in order. By setting $\lambda = 1$ we retrieve the case of a \emph{free-running} LO studied in, e.g.,~\cite{pollet95a, tomba98a, katz04a}. 
When $0<\lambda<1$, the process in~\eqref{eq:phase_noise} is stationary and corresponds to  a \emph{partially-coherent} LO. In this case, the process~\eqref{eq:phase_noise} models only residual phase errors after a phase-tracking scheme, e.g., a phase-locked loop (PLL)~\cite{petrovic07a}. 
We will focus on this case in the remainder of the paper.


\subsection{Analog-to-Digital Converter: Behavioral Model} \label{sec:adc}

The down-converted signal is further converted into the digital domain---a process that involves discretization in both time and amplitude. The in-phase and quadrature components of the down-converted signal $z_b(t)$ are converted separately using a pair of ADCs with $q$-bit resolution that operate at sampling rate $F_s$.
Let $z_b[n] = z_b(nT_s)$ denote the input to the quantizer in the ADCs of the $b$th BS antenna ($b = 0,1,\dots,B-1$) at time $nT_s$. 
Then, the corresponding ADC output can be written as
\begin{IEEEeqnarray}{rCl} \label{eq:f_adc}
	r_b[n] &=& f_\text{adc}(z_b[n]) = \quantize\lefto(\Re\{z_b[n]\}\right) + j 	\quantize\lefto(\Im\{z_b[n]\}\right) \IEEEeqnarraynumspace
\end{IEEEeqnarray}
where
\begin{IEEEeqnarray}{rCl} \label{eq:quantizer_uniform}
	Q(z) &=&
	\begin{cases}
		\frac{\Delta}{2}\lefto(1-2^q\right)  & \text{if $z < -\frac{\Delta}{2} 2^q$ } \\
		\Delta\lefto\lfloor \frac{z}{\Delta} \right\rfloor + \frac{\Delta}{2} & \text{if $\abs{z} < \frac{\Delta}{2} 2^q$} \\
		\frac{\Delta}{2}\lefto(2^q-1\right) & \text{if $z > \frac{\Delta}{2} 2^q$}
	\end{cases} \IEEEeqnarraynumspace
\end{IEEEeqnarray}
models a $q$-bit uniform midrise quantizer with step size $\Delta$. We assume that the  parameters of the behavioral model, i.e., the number of bits $q$ and the step size $\Delta$, are known to the~BS.

\subsection{Discrete-Time Channel Input-Output Relation}

Since both the channel input $\{ s_u[n] \}$ and output $\{ r_b[n] \}$ are discrete-time random processes, it is convenient to work with a discrete-time channel input-output relation.
To this end, let $\vecx[n] = [x_1[n], x_2[n], \dots, x_B[n]]^T$ and let $\vecr[n] = [r_1[n], r_2[n], \dots, r_B[n]]^T$. By inserting~\eqref{eq:f_lna} and~\eqref{eq:f_mixer} into \eqref{eq:f_adc} we can write the discrete-time channel input-output relation, including the BS-side nonlinearities, in matrix form as
\begin{IEEEeqnarray}{rCl} \label{eq:inout_vec_discrete}
\vecr[n]  &=& f_\text{adc}\lefto(  e^{j\phi[n]} f_\text{lna}\lefto( \vecx[n]\right) \right)
\end{IEEEeqnarray}
where the scalar functions $f_\text{adc}(\cdot)$ and $f_\text{lna}(\cdot)$ are applied element-wise to vector inputs.
Now, let $\vecs[n] = [s_1[n], s_2[n], \dots, s_U[n]]^T$ 
denote the time-domain symbols transmitted from the $U$ UEs at time instant $nT_s$. Then,
\begin{IEEEeqnarray}{rCl} \label{eq:inout_channel_vec_discrete}
	\vecx[n] &=& \sum_{\ell = 0}^{L-1} \matH[\ell]\vecs[n-\ell] + \vecw[n].
\end{IEEEeqnarray}
Here, the entry on the $b$th row and on the $u$th column of $\matH[\ell] \in \opC^{B \times U}$ is $h_{b,u}[\ell]$, where $h_{b,u}[\ell]$ denotes the $\ell$th tap of the discrete-time channel, which corresponds to the $\ell$th sample of the convolution between $h_{b,u}(\tau)$ and $\sinc(\tau/T_s)$. Furthermore, $\vecw[n] = [w_1[n], w_2[n], \dots, w_B[n]]^T \distas \jpg(\matzero_{B \times 1},N_0\matI_B)$ is the discrete-time AWGN with $w_b[n] = w_b(nT_s)$.

\section{Composite Hardware-Impairment Model} \label{sec:composite}
 
The nonlinearities $f_\text{lna}(\cdot)$ and $f_\text{adc}(\cdot)$ in the behavioral model~\eqref{eq:inout_vec_discrete} renders a theoretical performance analysis challenging.
We now derive an approximate, yet accurate, aggregate statistical model for the distortion caused by the hardware impairments reviewed in~\fref{sec:behave}.
Specifically, our approach consists of the following two steps. 
First, by assuming that the input to each nonlinear hardware component is Gaussian and by leveraging Bussgang's theorem~\cite{bussgang52a},  we  write the output of each nonlinear hardware component as a linear function of the input plus an additive distortion term that is uncorrelated with the input. 
Second, we combine the individual linearized models to obtain an aggregate linearized model for the joint distortion caused by the nonideal hardware components. 

It is worth clarifying at this point that, although the individual linearized models are exact, provided that the input to the corresponding nonlinear hardware component is Gaussian, the resulting aggregated model is only an approximation.
Indeed, although the input to the first nonlinear hardware component can  be assumed Gaussian if the transmitted signal is Gaussian, its output, which becomes the input to the next nonlinear hardware component, is typically not Gaussian.
Nevertheless, as we will demonstrate in~\fref{sec:numerical}, the Bussgang-based aggregate linearized model turns out to be surprisingly accurate.

\subsection{Linearization via Bussgang's Theorem}

We start by formalizing Bussgang's theorem.
Assume that $\{\vecp[n]\}$, where $\vecp[n]\in\opC^B$, is a stationary vector-valued circularly-symmetric Gaussian process and that $\vecq[n] = f(\vecp[n])$ where $f(\cdot)$ is some scalar-valued function that is applied element-wise on a vector.
Let $\matC_\vecp[m] = \Ex{}{\vecp[n]\vecp^H[n-m]}$   and let $\matC_{\vecq\vecp}[m] = \Ex{}{\vecq[n]\vecp^H[n-m]}$  Then, according to Bussgang's theorem~\cite{bussgang52a}, it holds~that
\begin{IEEEeqnarray}{rCl} \label{eq:bussgang}
	\matC_{\vecq\vecp}[m] &=& \matG\matC_{\vecp}[m]	
\end{IEEEeqnarray}
where the linear gain $\matG \in \opC^{B \times B}$ is a diagonal matrix with diagonal entries given by
\begin{IEEEeqnarray}{rCl} \label{eq:bussgang_gain}
\lefto[\matG\right]_{b,b} 
&=& \frac{1}{\Ex{}{|p_b[n]|^2}} \Ex{}{f(p_b[n])p_b^*[n]}.
\end{IEEEeqnarray}
It follows from~\eqref{eq:bussgang}~that
\begin{IEEEeqnarray}{rCl} \label{eq:bussgang_decomp_tot}
	\vecq[n] &=& \matG\vecp[n] + \vece[n]	
\end{IEEEeqnarray}
where the distortion $\{\vece[n]\}$, which is not Gaussian distributed in general, is uncorrelated with $\{\vecp[n]\}$. 
%
%
The autocovariance of the distortion~$\matC_\vece[m] = \Ex{}{\vece[n]\vece^H[n-m]}$ is readily obtained from~\eqref{eq:bussgang_decomp_tot} as follows:
\begin{IEEEeqnarray}{rCl} \label{eq:bussgang_dist}
	\matC_\vece[m] &=& \matC_\vecq[m] - \matG \matC_\vecp[m] \matG^H.	
\end{IEEEeqnarray}

In what follows, we  use Bussgang's theorem to linearize the input-output relation of each nonlinear hardware component separately. 
We then combine the resulting linearized expressions to obtain a simple, yet accurate, approximation for the ADC output $\vecr[n]$ in~\eqref{eq:inout_vec_discrete} as a function of the LNA input in~\eqref{eq:f_lna}.

Starting from the LNA, if we assume $\vecx[n]$ in~\eqref{eq:inout_channel_vec_discrete} to be circularly symmetric Gaussian distributed, we can write the LNA output $\vecy[n] = [y_1[n], y_2[n], \dots, y_B[n]]^T$ as
\begin{IEEEeqnarray}{rCl} 
\vecy[n] &= f_\text{lna}\lefto( \vecx[n] \right) &= \matG_\text{lna}\vecx[n] + \vece_\text{lna}[n]. \label{eq:bussgang_decomp_lna}
\end{IEEEeqnarray}
Note that $\vecx[n]$ is Gaussian if we take the input symbols to be Gaussian, i.e., $\vecs[n]\distas\jpg(\veczero_{U \times 1}, \frac{S}{N} \matI_U)$ and the channel $\{\matH[\ell]\}$ to be known at the BS. 
We will adopt these two assumptions in the remainder of the paper.

Similarly, if we assume $\{\vecy[n]\}$ to be a circularly-symmetric stationary Gaussian process, we can write the down-converted signal $\vecz[n] = [z_1[n], z_2[n], \dots, z_B[n]]^T$ as
\begin{IEEEeqnarray}{rCl} 
\vecz[n] &= e^{j\phi[n]}\vecy[n] &= \matG_\text{osc}\vecy[n] + \vece_\text{osc}[n]. \label{eq:bussgang_decomp_osc}
\end{IEEEeqnarray}
Finally, if we let $\{\vecz[n]\}$ be a circularly-symmetric stationary Gaussian process, we can write the ADC output $\vecr[n]$ as
\begin{IEEEeqnarray}{rCl} 
\vecr[n] &= f_\text{adc}\lefto( \vecz[n] \right) &= \matG_\text{adc}\vecz[n] + \vece_\text{adc}[n]. \label{eq:bussgang_decomp_adc}	
\end{IEEEeqnarray}

Next, we combine~\eqref{eq:bussgang_decomp_lna}, \eqref{eq:bussgang_decomp_osc}, and \eqref{eq:bussgang_decomp_adc}	 to obtain the following approximation for~$\vecr[n]$ in~\eqref{eq:inout_vec_discrete}:
\begin{IEEEeqnarray}{rCl}
\vecr[n]
&\approx& \matG_\text{adc}\matG_\text{osc}\matG_\text{lna}\vecx[n] + \vece_\text{adc}[n]  \nonumber\\
&& +\,  \matG_\text{adc}\vece_\text{osc}[n] + \matG_\text{adc}\matG_\text{osc}\vece_\text{lna}[n] \label{eq:bussgang_approx_1} \\
&=& \matG_\text{tot}\vecx[n] + \vece_\text{tot}[n] \label{eq:bussgang_approx_2}.
\end{IEEEeqnarray}
Here, $\vece_\text{tot}[n] = \vece_\text{adc}[n] + \matG_\text{adc}\vece_\text{osc}[n] + \matG_\text{adc}\matG_\text{osc}\vece_\text{lna}[n]$ and $\matG_\text{tot} = \matG_\text{adc}\matG_\text{osc}\matG_\text{lna}$.
We emphasize that~\eqref{eq:bussgang_approx_1} is an approximation since the output of the first nonlinear hardware component in our chain, i.e., the LNA, is not Gaussian distributed. Nevertheless, as we will demonstrate in~\fref{sec:numerical}, the approximation~\eqref{eq:bussgang_approx_1} turns out to be surprisingly accurate.

The frequency-domain received hardware-impaired vector on the $k$th subcarrier ($k = 0,1,\dots,N-1$) is given by
\begin{IEEEeqnarray}{rCl} \label{eq:rk_vec}
	\hat\vecr[k] &=& \frac{1}{\sqrt{N}} \sum_{n = 0}^{N-1}  \vecr[n]  e^{-jk\frac{2\pi}{N}n}.
\end{IEEEeqnarray}
Let $\hat\vecs[k] = [\hat{s}_1, \hat{s}_2, \dots, \hat{s}_{U}]^T$. Then, by substituting~\eqref{eq:inout_channel_vec_discrete} and~\eqref{eq:bussgang_approx_2} into~\eqref{eq:rk_vec} we obtain the following approximation for $\hat\vecr[k]$:
\begin{IEEEeqnarray}{rCl} \label{eq:bussgang_approx_freq}
\hat\vecr[k] 
&\approx& \matG_\text{tot} \lefto(\widehat\matH[k]\hat\vecs[k] + \hat\vecw[k]\right) + \hat\vece_\text{tot}[k]. \IEEEeqnarraynumspace
\end{IEEEeqnarray}
Here,~$\widehat\matH[k] = \sum_{\ell = 0}^{L-1}\matH[\ell]e^{-jk\frac{2\pi}{N}\ell}$ is the frequency-domain channel matrix associated with the $k$th subcarrier. Furthermore, $\hat\vece_\text{tot}[k] = \frac{1}{\sqrt{N}} \sum_{n = 0}^{N-1} \vece_\text{tot}[n] e^{-jk\frac{2\pi}{N}n}$ and $\hat\vecw[k] =  \frac{1}{\sqrt{N}}\sum_{n = 0}^{N-1} \vecw[n] e^{-jk\frac{2\pi}{N}n} \distas \jpg(\matzero_{B \times 1}, N_0\matI_B)$ is the frequency-domain distortion and AWGN, respectively.
From~\eqref{eq:bussgang_approx_freq}, we find that the covariance $\matC_{\hat\vecr}[k] = \Ex{}{\hat\vecr[k]\hat\vecr^H[k]}$ of $\hat\vecr[k]$ for $k = 0,1,\dots,N-1$ can be approximated as
\begin{IEEEeqnarray}{rCl} \label{eq:Cr_k}
\matC_{\hat\vecr}[k] 
&\approx& \matG_\text{tot} \matC_{\hat\vecx}[k] \matG_\text{tot}^H + \matC_{\hat\vece_\text{tot}}[k] \IEEEeqnarraynumspace
\end{IEEEeqnarray}
where $\matC_{\hat\vecx}[k] = \opE\big[\hat\vecx[k]\hat\vecx^H[k]\big] =\widehat\matH[k]\matC_{\hat\vecs}[k]\widehat\matH^H[k] + N_0\matI_B$.
Here, $\matC_{\hat\vecs}[k] = \opE\big[\hat\vecs[k]\hat\vecs^H[k]\big]$ is the covariance of $\hat\vecs[k]$, which is equal to $\matI_B$ if $k \in \setS$ and to $\matzero_{B \times B}$ if $k \in \setG$. Furthermore,
\begin{IEEEeqnarray}{rCl} \label{eq:Ce_tot_k}
\matC_{\hat\vece_\text{tot}}[k] = \Ex{}{\hat\vece_\text{tot}[k]\hat\vece_\text{tot}^H[k]} = \sum_{m=0}^{N-1} \matC_{\vece_\text{tot}}[m] e^{-jk\frac{2\pi}{N}m} \IEEEeqnarraynumspace
\end{IEEEeqnarray}
where 
\begin{IEEEeqnarray}{rCl} \label{eq:Ce_tot_time}
\matC_{\vece_\text{tot}}[m] 
&=& \Ex{}{\vece_\text{tot}[n]\vece_\text{tot}^H[n-m]} \\
&=& \matC_{\vece_\text{adc}}[m] + \matG_\text{adc}\matC_{\vece_\text{osc}}[m]\matG_\text{adc}^H \nonumber\\
&&+\, \matG_\text{osc}\matG_\text{adc}\matC_{\vece_\text{osc}}[m]\matG_\text{adc}^H\matG_\text{osc}^H.
\end{IEEEeqnarray}
The covariance matrices~\eqref{eq:Cr_k} and~\eqref{eq:Ce_tot_k} can be used to analyze the PSD of the hardware-impaired signal. Indeed, they describe how the per-subcarrier received signal and the corresponding distortion are correlated in the spatial domain.
They also allow us to separate the contributions due to the intended signal from the one due to  distortion, which is useful in performance analyses, as exemplified in~\fref{sec:ber}. 
We now provide closed-form expressions for $\matG_\text{lna}$, $\matG_\text{osc}$, and $\matG_\text{adc}$, as well as for the autocovariance of the three distortion processes $\{\vece_\text{lna}[n]\}$, $\{\vece_\text{osc}[n]\}$, and~$\{\vece_\text{adc}[n]\}$, which are all required for evaluating~\eqref{eq:Cr_k} and~\eqref{eq:Ce_tot_k}.

\subsection{Low-Noise Amplifier: Linearized Model} 

By inserting the behavioral model~\eqref{eq:f_lna} into \eqref{eq:bussgang_gain} we find that $\lefto[\matG_\text{lna}\right]_{b,b} =  \alpha_1 + 2\alpha_2\Ex{}{|x_b[n]|^2}$ for $b = 1, 2, \dots, B$. 
Hence,
\begin{IEEEeqnarray}{rCl} \label{eq:G_lna}
\matG_\text{lna} &=& \alpha_1\matI_B + 2\alpha_2\diag\lefto( \matC_\vecx[0]\right).
\end{IEEEeqnarray}
Similarly, we have that
\begin{IEEEeqnarray}{rCl}
	\matC_\vecy[m] 
	&=& \Ex{}{\vecy[n]\vecy^H[n-m]} \\
	&=& \Ex{}{f_\text{lna}(\vecx[n])f_\text{lna}(\vecx[n-m])^H} \\
	&=& \abs{\alpha_1}^2\matC_\vecx[m] + 2\abs{\alpha_2}^2 \matC_\vecx[m] \circ \abs{\matC_\vecx[m]}^2 \IEEEeqnarraynumspace\nonumber\\
	&& + 2 \alpha_1^*\alpha_2 \diag\lefto( \matC_\vecx[0]\right) \matC_\vecx[m] \nonumber\\
	&& + 2 \alpha_2^*\alpha_1 \matC_\vecx[m] \diag\lefto( \matC_\vecx[0]\right) \nonumber\\
	&& + 2\abs{\alpha_2}^2 \diag\lefto( \matC_\vecx[0]\right) \matC_\vecx \diag\lefto( \matC_\vecx[0]\right). \label{eq:Cy} \IEEEeqnarraynumspace
\end{IEEEeqnarray}
Here,~$\matC_\vecx[m] \!=\! \Ex{}{\vecx[n]\vecx^H[n\!-\!m]} \!= \!\frac{1}{N}\sum_{k=0}^{N-1} \matC_{\hat\vecx}[k]e^{jk\frac{2\pi}{N}m}$.
Let now $\matC_{\vece_\text{lna}}[m] = \Ex{}{\vece_\text{lna}[n]\vece_\text{lna}^H[n-m]}$. By substituting \eqref{eq:G_lna} and \eqref{eq:Cy} into \eqref{eq:bussgang_dist}, we find that 
\begin{IEEEeqnarray}{rCl} \label{eq:Ce_lna}
\matC_{\vece_\text{lna}}	[m] &=& 2 \abs{\alpha_2}^2 \matC_\vecx[m] \circ \abs{\matC_\vecx[m]}^2.
\end{IEEEeqnarray}

\subsection{Phase Noise: Linearized Model}

By inserting~\eqref{eq:f_mixer} into \eqref{eq:bussgang_gain} we find that $\lefto[ \matG_\text{osc}\right]_{b,b} = \opE\big[e^{j\phi[n]}\big]$ for $b = 1,2,\dots,B$. Since we assumed that $0 < \lambda < 1$, the random process $\{ \phi[n] \}$ in~\eqref{eq:phase_noise} is a stationary Gaussian process with zero mean and~variance $\sigma_\phi^2 = \Ex{}{\phi^2[n]} = \frac{2\pi\beta T_s}{1-\lambda^2}$.
Hence, $\matG_\text{osc}$ in~\eqref{eq:bussgang_decomp_osc} is given by 
\begin{IEEEeqnarray}{rCl} \label{eq:G_osc}
	\matG_\text{osc} &=& \opE\big[e^{j\phi[n]}\big]	\matI_B = e^{-\frac{1}{2}\sigma_\phi^2}\matI_B= e^{-\frac{\pi\beta T_s}{1-\lambda^2}}\matI_B.
\end{IEEEeqnarray}
Here, we used that $\opE\big[e^{jv\phi[n]}\big] = e^{-\frac{1}{2}v^2\sigma_\phi^2}$ is the characteristic function of $\phi[n] \distas \normal(0, \sigma_\phi^2)$. 
%
%
Similarly,
\begin{IEEEeqnarray}{rCl}
\matC_\vecz[m]
&=& \Ex{}{\vecz[n]\vecz^H[n-m]}	\\
&=& \opE\big[e^{j(\phi[n] - \phi[n-m])}\big]\Ex{}{\vecy[n]\vecy^H[n-m]} \\
&=& e^{- \sigma_\phi^2 +  \rho_\phi(m)} \matC_\vecy[m] \label{eq:Cz}
\end{IEEEeqnarray}
where $\rho_\phi(m) = \Ex{}{\phi[n]\phi[n-m]} = \frac{2\pi\beta T_s}{1-\lambda^2}\lambda^{\abs{m}}$.
%
%
Let now $\matC_{\vece_\text{osc}}[m] = \Ex{}{\vece_\text{osc}[n]\vece_\text{osc}^H[n-m]}$. By substituting~\eqref{eq:G_osc} and~\eqref{eq:Cz} into~\eqref{eq:bussgang_dist}, we find~that 
\begin{IEEEeqnarray}{rCl} \label{eq:Ce_lna}
\matC_{\vece_\text{osc}}[m] 
&=& \lefto(e^{-\frac{2\pi\beta T_s}{1-\lambda^2}\lefto( 1 - \lambda^{\abs{m}} \right)} -  e^{-\frac{2\pi\beta T_s}{1-\lambda^2}} \right)\matC_{\vecy}[m].
\end{IEEEeqnarray}

\subsection{Analog-to-Digital Converter: Linearized Model}

Many recent papers have used Bussgang's theorem to analyze the distortion caused by amplitude quantization (see, e.g.,~\cite{jacobsson17b, mollen16c, zhang12a, li17b, jacobsson17d, jacobsson17e,mezghani12b}).
As pointed out in these contributions (see, e.g.,~\cite[Eq.~(14)]{jacobsson17d}), the matrix  $\matG_\text{adc}$ in~\eqref{eq:bussgang_decomp_adc} is given by
\begin{IEEEeqnarray}{rCl} \label{eq:G_adc}
\matG_\text{adc} &=& \frac{\Delta}{\sqrt{\pi}} \, \diag\lefto(\matC_\vecz[0]\right)^{-1/2} \nonumber \\
&&\times\sum_{i=1}^{2^q-1} \exp\lefto(-\Delta^2\lefto( i - 2^{q-1} \right)^2 \diag\lefto(\matC_\vecz[0]\right)^{-1}\right). \IEEEeqnarraynumspace
\end{IEEEeqnarray}
Let now $\matC_{\vece_\text{adc}}[m] = \Ex{}{\vece_\text{adc}[n]\vece_\text{adc}^H[n-m]}$. Unfortunately, no closed-form expression for $\matC_{\vece_\text{adc}}[m]$ is available for $q > 1$. Therefore, in what follows, we  evaluate $\matC_{\vece_\text{adc}}[m]$ using the diagonal approximation put forward in~\cite[Sec.~IV]{jacobsson17e}. Specifically, we approximate $\matC_{\vece_\text{adc}}[m]$ by
\begin{IEEEeqnarray}{rCl} \label{eq:diagonal_approx}
\matC_{\vece_\text{adc}}[0]	&\approx& \frac{\Delta^2}{2}\lefto( 2^q - 1 \right)^2\matI_B  -\matG_\text{adc}\diag\lefto(\matC_{\vecz}[0]\right)\!\matG_\text{adc}  \nonumber\\
&& -4\Delta^2 \!\sum_{i=1}^{2^q-1} \! \lefto(i - 2^{q-1}\right) \nonumber\\
&& \times\lefto(1- Q\lefto( \sqrt{2}\lefto(i - 2^{q-1}\right)\diag\lefto(\matC_\vecz[0]\right)^{-1/2} \right)\right) \IEEEeqnarraynumspace
\end{IEEEeqnarray}
for $m=0$ and by $\matC_{\vece_\text{adc}}[m] = \matzero_{B \times B}$ for $m \neq 0$.
Here, $Q(x) = \frac{1}{\sqrt{2\pi}}\int_{x}^\infty e^{-u^2/2} \, \fack{\!u}$.
As shown in \cite[Sec.~V]{jacobsson17e}, this approximation is accurate for ADCs with $q\geq 3$. 

\newcommand{\figscale}{0.9}
\newcommand{\figsep}{\hspace{1.0cm}}
\begin{figure*}[!t]
\centering
\subfloat[PSD of $\{\vecr{[n]}\}$ when the only nonideal hardware component at the BS is the LNA; $\alpha_1 = 1.065$ and $\alpha_2 = -0.028$.]{\includegraphics[width=\figscale\columnwidth]{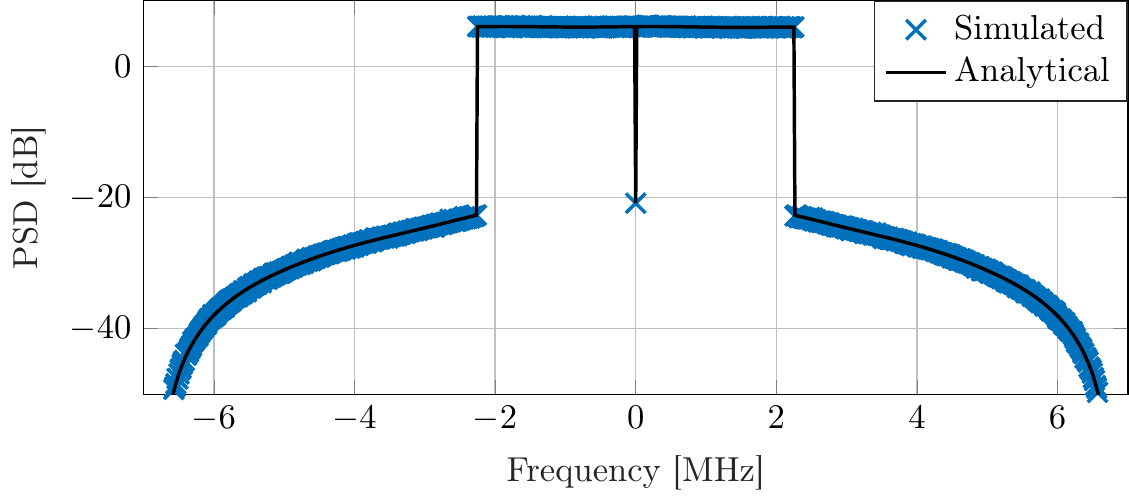}\label{fig:psd_amp}}	
\figsep
\subfloat[PSD of $\{\vecr{[n]}\}$ when the only nonideal hardware component at the BS is the LO; $\lambda = 0.99$ and $\beta = 10^3$.]{\includegraphics[width=\figscale\columnwidth]{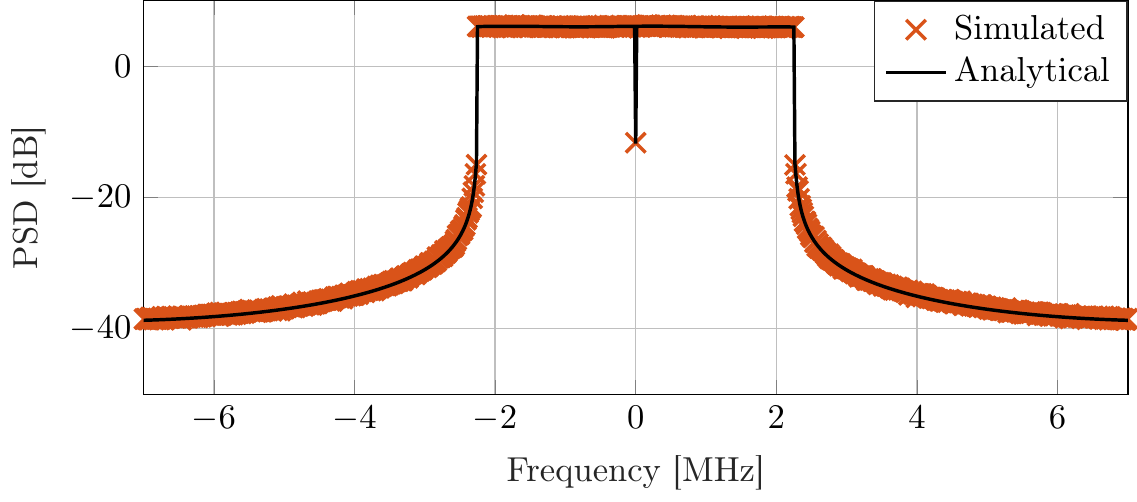}\label{fig:psd_osc}}	\\
\subfloat[PSD of $\{\vecr{[n]}\}$ when the only nonideal hardware component at the BS is the ADC; $q = 6$ and $\Delta = 0.086\sqrt{1.172 + N_0}$.]{\includegraphics[width=\figscale\columnwidth]{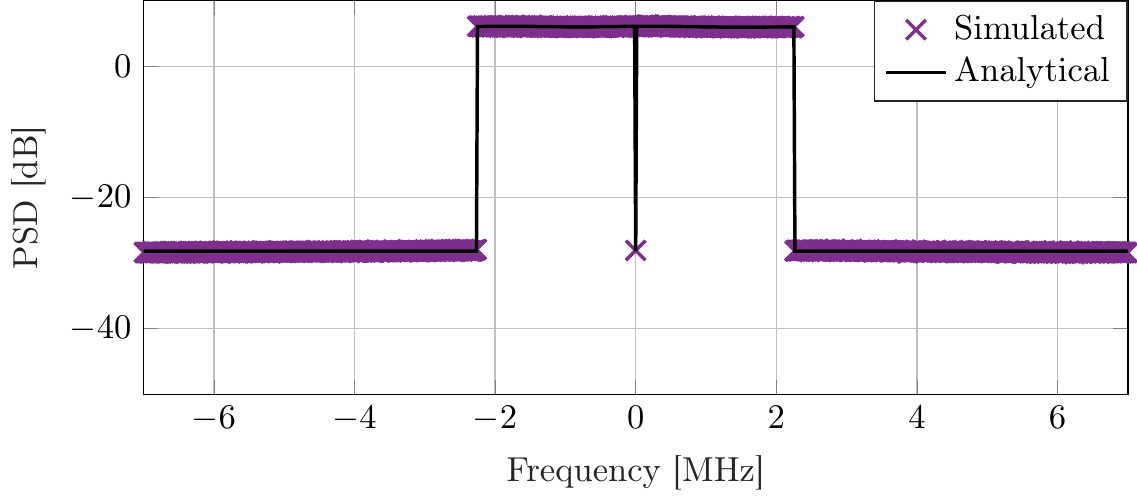}\label{fig:psd_adc}}	
\figsep	
\subfloat[PSD of $\{\vecr{[n]}\}$ when all hardware components, i.e., the LNAs, the LO, and the ADCs, are nonideal; parameters taken from (a), (b), and (c).]{\includegraphics[width=\figscale\columnwidth]{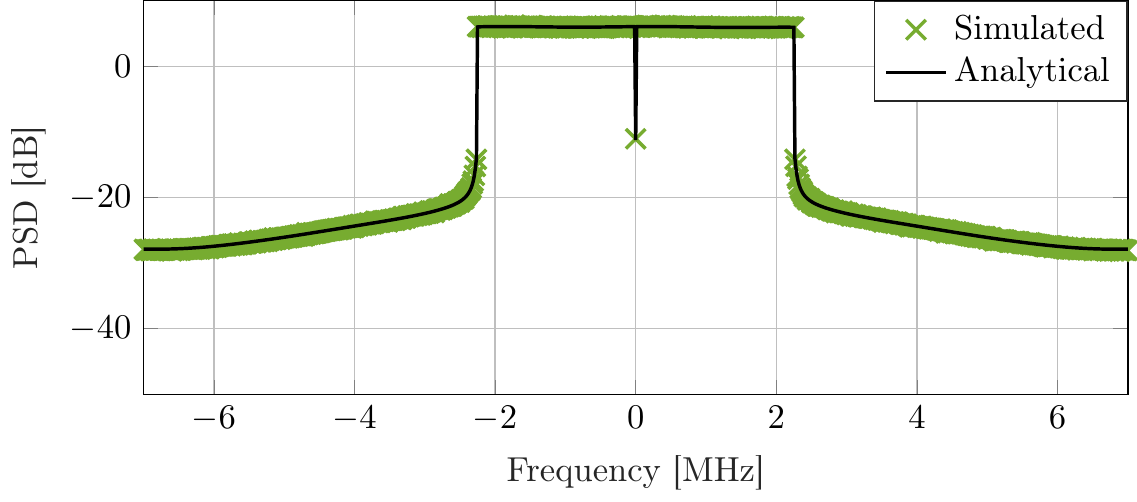}\label{fig:psd_all}}	
\vspace{0.05cm}
\caption{PSD of $\{\vecr[n]\}$; $B = 32$, $U = 4$, $L=10$, $S = 300$, $N = 1024$, and $N_0 = 0$. The colored markers correspond to numerical simulations; the black lines correspond to analytical results. 
Note that the analytical PSDs in \fref{fig:psd_amp} and \fref{fig:psd_osc} are exact for Gaussian inputs. 
The analytical PSD in \fref{fig:psd_adc} verifies the accuracy of the approximation~\eqref{eq:diagonal_approx}. The analytical PSD in \fref{fig:psd_all} verifies the accuracy of the composite hardware-impairment model presented in~\fref{sec:composite}.}
\label{fig:psd}
\end{figure*}

\section{Numerical Results} \label{sec:numerical}

We now demonstrate the accuracy of our models and their usefulness in performance analyses. 
In what follows, the number of BS antennas is $B = 32$ and the number of UEs is $U=4$. We consider LTE-inspired OFDM parameters. Specifically, the occupied subcarriers are the first $S/2$ to the left and the first $S/2$ to right of the DC subcarrier, i.e., $\setS = \{ 1, 2, \dots, S/2\} \union \{ N-S/2, N-S/2+1, \dots, N-1\}$.
Furthermore, the number of occupied subcarriers is $S = 300$, the total number of subcarriers is $N = 1024$, and the subcarrier spacing is $F_\text{sub} = 15$~kHz. Hence, the sampling rate is $F_s = NF_\text{sub} = 15.36$~MHz and the OSR is $3.41$.
We set the polynomial coefficients of the LNA to $\alpha_1 = 1.065$ and $\alpha_2 = -0.028$. For the phase-noise process, we set $\lambda = 0.99$ and $\beta = 10^3$.
For the ADCs, the number of bits is $q=6$ and the step size is $\Delta = 0.086\sqrt{1.172 + N_0}$.
We draw the elements of $\matH[\ell]$ for $\ell = 0,1,\dots,L-1$ independently from the $\jpg(0, L^{-1})$ distribution.
The number of nonzero channel taps is~$L = 10$. 

For the numerical simulations, we draw each transmitted symbol~$\hat\vecs[k]$ for $k \in \setS$ independently from a quadrature phase-shift keying~(QPSK) constellation. 
Although our theoretical analysis requires the received time-domain signal to be Gaussian, drawing the input symbols from a QPSK constellation leads to negligible errors. 
Indeed, the received time-domain signal is the sum of $US = 1200$ independent and identically distributed random variables, which can be well-approximated by a Gaussian random variable according to the central limit~theorem.

\subsection{Power Spectral Density}

In~\fref{fig:psd}, we plot the PSD of~$\{\vecr[n]\}$.
The average value of the PSD on the $k$th subcarrier, averaged over the $B$ BS antennas, is obtained analytically by computing~$\frac{1}{B}\vecnorm{\diag\lefto(\matC_{\hat\vecr}[k]\right)}^2_2$ for $k = 0,1,\dots,N-1$, where $\matC_{\hat\vecr}[k]$ is given in~\eqref{eq:Cr_k}. 
The simulated curve is obtained via Monte-Carlo sampling.
We note from \fref{fig:psd_all} that the PSD obtained using the aggregate hardware-impairment model~\eqref{eq:Cr_k} describes accurately the spectral regrowth caused by the nonideal hardware components and agrees perfectly with the numerical simulations, confirming the accuracy of our linearized aggregate model.

\begin{figure}[!t]
\centering
\includegraphics[width=.9\columnwidth]{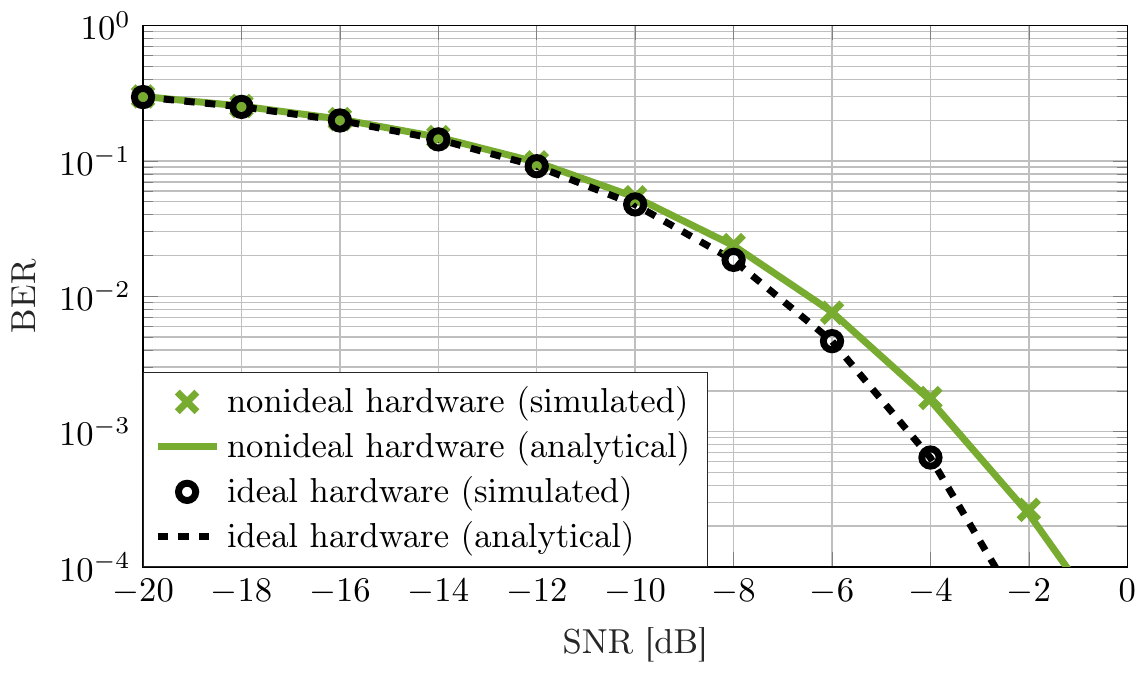}
\caption{Uncoded BER with zero forcing and QPSK; $B = 32$, $U = 4$, $L=10$, $S = 300$, $N = 1024$, $\alpha_1 = 1.065$, $\alpha_2 = -0.028$, $\beta = 10^3$, $\lambda = 0.99$, $q = 6$, and $\Delta = 0.086\sqrt{1.172 + N_0}$. Here, $\textit{SNR} = 1/N_0$. The markers correspond to simulation results and the lines to the analytical approximation proposed in~\eqref{eq:ber}. Clearly, our analysis is extremely accurate.}
\label{fig:ber}
\end{figure}

\subsection{Bit-Error Rate} \label{sec:ber} 

A zero-forcing estimate $\hat\vecs_\text{est}[k]$ of the transmitted symbols $\hat\vecs[k]$ is obtained from $\hat\vecr[k]$ for $k = 0,1,\dots,N-1$~as 
\begin{IEEEeqnarray}{rCl} \label{eq:forza_zero}
	\hat\vecs_\text{est}[k] &=& \widehat\matA^H[k] \,\hat\vecr[k]
\end{IEEEeqnarray}
where $\widehat\matA[k] = \matG_\text{tot}\widehat\matH[k]\big( \matG_\text{tot}\widehat\matH[k]\widehat\matH^H[k]\matG_\text{tot}^H\big)^{-1}$.
By inserting the approximation~\eqref{eq:bussgang_approx_freq} into \eqref{eq:forza_zero}, we find a linear relationship between $\hat{\vecs}[k]$ and $\hat{\vecs}_\text{est}[k]$, which makes it straightforward to compute the resulting signal-to-interference-noise-and-distortion ratio (SINDR) for the $u$th UE ($u = 1,2,\dots,U$) as follows:
\begin{IEEEeqnarray}{rCl} \label{eq:sindr}
\textit{SINDR}_{u}[k] 
&\approx& \frac{\big\lvert\widehat\veca_{u}^H[k] \matG_\text{tot} \hat\vech_{u}[k]\big\rvert^2 }{  I_{u}[k] + N_{u}[k] + D_{u}[k]}.
\end{IEEEeqnarray}
Here, $\hat\veca_u[k] \in \opC^B$ is the $u$th column of $\widehat\matA[k]$, $\hat\vech_u[k] \in \opC^B$ is the $u$th column of $\widehat\matH[k]$, $I_{u}[k] = \sum_{v \neq u}\big\lvert\hat\veca_{u}^H[k]\matG_\text{tot} \hat\vech_{v}[k]\big\rvert^2$, $N_{u}[k] = N_0 \|\hat\veca_{u}^H[k]\matG_\text{tot}\|_2^2$, and $D_{u}[k] = \hat\veca_{u}^H[k]\matC_{\hat\vece_\text{tot}}[k] \,\hat\veca_{u}[k]$.

In~\fref{fig:ber}, we show the uncoded BER with zero forcing and QPSK signaling, which we approximate as
\begin{IEEEeqnarray}{rCl} \label{eq:ber}
\textit{BER} &\approx& \frac{1}{US}\sum_{u=1}^U \sum_{k \in \setS} Q\lefto( \sqrt{\textit{SINDR}_u[k]}\right)
\end{IEEEeqnarray}
where $\textit{SINDR}_u[k]$ is given in~\eqref{eq:sindr}.
In~\fref{fig:ber}, we have averaged the BER over $100$ channel realizations. We note that the performance predicted by the composite hardware-impairment model is again in excellent agreement with the numerical results.

\section{Conclusions and Future Work}

In this work, we have presented a linearized aggregate model for BS-side hardware impairments, which is derived from behavioral models of nonideal hardware components via Bussgang's decomposition.
Although the proposed aggregate model is not exact because it relies on the Gaussianity of the input to each nonlinear hardware component (which does, in general, not hold in practice), it turns out to be extremely accurate for many scenarios of interest, such as the massive MU-MIMO-OFDM uplink system considered in~\fref{sec:numerical}.

The analysis presented in this paper can be extended to the case of pilot-based channel estimation and to the MU-MIMO-OFDM downlink.
One limitation of our current framework is that it is confined to memoryless nonlinearities  and does not capture other sources of hardware impairments, such as in-phase/quadrature imbalance and mutual coupling.
An extension of our analysis to incorporate these impairments  is part of ongoing work.

%
%
%
%

\balance

\bibliographystyle{IEEEtran}
\bibliography{IEEEabrv,confs-jrnls,publishers,svenbib}

\end{document}

%% file: vmr-symbols-vecbold.tex
\usepackage{amssymb}
\usepackage{amsfonts}
\usepackage{mathrsfs}
\usepackage{xspace}
\usepackage{bm}
\usepackage{upgreek}

\newcommand{\safemath}[2]{\newcommand{#1}{\ensuremath{#2}\xspace}}

\safemath{\bma}{\mathbf{a}}
\safemath{\bmb}{\mathbf{b}}
\safemath{\bmc}{\mathbf{c}}
\safemath{\bmd}{\mathbf{d}}
\safemath{\bme}{\mathbf{e}}
\safemath{\bmf}{\mathbf{f}}
\safemath{\bmg}{\mathbf{g}}
\safemath{\bmh}{\mathbf{h}}
\safemath{\bmi}{\mathbf{i}}
\safemath{\bmj}{\mathbf{j}}
\safemath{\bmk}{\mathbf{k}}
\safemath{\bml}{\mathbf{l}}
\safemath{\bmm}{\mathbf{m}}
\safemath{\bmn}{\mathbf{n}}
\safemath{\bmo}{\mathbf{o}}
\safemath{\bmp}{\mathbf{p}}
\safemath{\bmq}{\mathbf{q}}
\safemath{\bmr}{\mathbf{r}}
\safemath{\bms}{\mathbf{s}}
\safemath{\bmt}{\mathbf{t}}
\safemath{\bmu}{\mathbf{u}}
\safemath{\bmv}{\mathbf{v}}
\safemath{\bmw}{\mathbf{w}}
\safemath{\bmx}{\mathbf{x}}
\safemath{\bmy}{\mathbf{y}}
\safemath{\bmz}{\mathbf{z}}
\safemath{\bmzero}{\mathbf{0}}
\safemath{\bmone}{\mathbf{1}}

\bmdefine{\biad}{a}
\bmdefine{\bibd}{b}
\bmdefine{\bicd}{c}
\bmdefine{\bidd}{d}
\bmdefine{\bied}{e}
\bmdefine{\bifd}{f}
\bmdefine{\bigd}{g}
\bmdefine{\bihd}{h}
\bmdefine{\biid}{i}
\bmdefine{\bijd}{j}
\bmdefine{\bikd}{k}
\bmdefine{\bild}{l}
\bmdefine{\bimd}{m}
\bmdefine{\bind}{n}
\bmdefine{\biod}{o}
\bmdefine{\bipd}{p}
\bmdefine{\biqd}{q}
\bmdefine{\bird}{r}
\bmdefine{\bisd}{s}
\bmdefine{\bitd}{t}
\bmdefine{\biud}{u}
\bmdefine{\bivd}{v}
\bmdefine{\biwd}{w}
\bmdefine{\bixd}{x}
\bmdefine{\biyd}{y}
\bmdefine{\bizd}{z}

\bmdefine{\bixid}{\xi}
\bmdefine{\bilambdad}{\lambda}
\bmdefine{\bimud}{\mu}
\bmdefine{\bithetad}{\theta}
\bmdefine{\biphid}{\phi}
\bmdefine{\bideltad}{\delta}

\safemath{\bmia}{\biad}
\safemath{\bmib}{\bibd}
\safemath{\bmic}{\bicd}
\safemath{\bmid}{\bidd}
\safemath{\bmie}{\bied}
\safemath{\bmif}{\bifd}
\safemath{\bmig}{\bigd}
\safemath{\bmih}{\bihd}
\safemath{\bmii}{\biid}
\safemath{\bmij}{\bijd}
\safemath{\bmik}{\bikd}
\safemath{\bmil}{\bild}
\safemath{\bmim}{\bimd}
\safemath{\bmin}{\bind}
\safemath{\bmio}{\biod}
\safemath{\bmip}{\bipd}
\safemath{\bmiq}{\biqd}
\safemath{\bmir}{\bird}
\safemath{\bmis}{\bisd}
\safemath{\bmit}{\bitd}
\safemath{\bmiu}{\biud}
\safemath{\bmiv}{\bivd}
\safemath{\bmiw}{\biwd}
\safemath{\bmix}{\bixd}
\safemath{\bmiy}{\biyd}
\safemath{\bmiz}{\bizd}

\safemath{\bmxi}{\bixid}
\safemath{\bmlambda}{\bilambdad}
\safemath{\bmmu}{\bimud}
\safemath{\bmtheta}{\bithetad}
\safemath{\bmphi}{\biphid}
\safemath{\bmdelta}{\bideltad}

\safemath{\bA}{\mathbf{A}}
\safemath{\bB}{\mathbf{B}}
\safemath{\bC}{\mathbf{C}}
\safemath{\bD}{\mathbf{D}}
\safemath{\bE}{\mathbf{E}}
\safemath{\bF}{\mathbf{F}}
\safemath{\bG}{\mathbf{G}}
\safemath{\bH}{\mathbf{H}}
\safemath{\bI}{\mathbf{I}}
\safemath{\bJ}{\mathbf{J}}
\safemath{\bK}{\mathbf{K}}
\safemath{\bL}{\mathbf{L}}
\safemath{\bM}{\mathbf{M}}
\safemath{\bN}{\mathbf{N}}
\safemath{\bO}{\mathbf{O}}
\safemath{\bP}{\mathbf{P}}
\safemath{\bQ}{\mathbf{Q}}
\safemath{\bR}{\mathbf{R}}
\safemath{\bS}{\mathbf{S}}
\safemath{\bT}{\mathbf{T}}
\safemath{\bU}{\mathbf{U}}
\safemath{\bV}{\mathbf{V}}
\safemath{\bW}{\mathbf{W}}
\safemath{\bX}{\mathbf{X}}
\safemath{\bY}{\mathbf{Y}}
\safemath{\bZ}{\mathbf{Z}}

\safemath{\bZero}{\mathbf{0}}
\safemath{\bOne}{\mathbf{1}}
\safemath{\bDelta}{\mathbf{\Delta}}
\safemath{\bLambda}{\mathbf{\UpLambda}}
\safemath{\bPhi}{\mathbf{\Upphi}}
\safemath{\bSigma}{\mathbf{\Upsigma}}
\safemath{\bOmega}{\mathbf{\Upomega}}
\safemath{\bTheta}{\mathbf{\Uptheta}}

\bmdefine{\biAd}{A}
\bmdefine{\biBd}{B}
\bmdefine{\biCd}{C}
\bmdefine{\biDd}{D}
\bmdefine{\biEd}{E}
\bmdefine{\biFd}{F}
\bmdefine{\biGd}{G}
\bmdefine{\biHd}{H}
\bmdefine{\biId}{I}
\bmdefine{\biJd}{J}
\bmdefine{\biKd}{K}
\bmdefine{\biLd}{L}
\bmdefine{\biMd}{M}
\bmdefine{\biOd}{N}
\bmdefine{\biPd}{O}
\bmdefine{\biQd}{P}
\bmdefine{\biRd}{R}
\bmdefine{\biSd}{S}
\bmdefine{\biTd}{T}
\bmdefine{\biUd}{U}
\bmdefine{\biVd}{V}
\bmdefine{\biWd}{W}
\bmdefine{\biXd}{X}
\bmdefine{\biYd}{Y}
\bmdefine{\biZd}{Z}

\bmdefine{\biDelta}{\Delta}
\bmdefine{\biLambda}{\Lambda}
\bmdefine{\biPhi}{\Phi}
\bmdefine{\biSigma}{\Sigma}
\bmdefine{\biOmega}{\Omega}
\bmdefine{\biTheta}{\Theta}

\safemath{\bimA}{\biAd}
\safemath{\bimB}{\biBd}
\safemath{\bimC}{\biCd}
\safemath{\bimD}{\biDd}
\safemath{\bimE}{\biEd}
\safemath{\bimF}{\biFd}
\safemath{\bimG}{\biGd}
\safemath{\bimH}{\biHd}
\safemath{\bimI}{\biId}
\safemath{\bimJ}{\biJd}
\safemath{\bimK}{\biKd}
\safemath{\bimL}{\biLd}
\safemath{\bimM}{\biMd}
\safemath{\bimN}{\biNd}
\safemath{\bimO}{\biOd}
\safemath{\bimP}{\biPd}
\safemath{\bimQ}{\biQd}
\safemath{\bimR}{\biRd}
\safemath{\bimS}{\biSd}
\safemath{\bimT}{\biTd}
\safemath{\bimU}{\biUd}
\safemath{\bimV}{\biVd}
\safemath{\bimW}{\biWd}
\safemath{\bimX}{\biXd}
\safemath{\bimY}{\biYd}
\safemath{\bimZ}{\biZd}

\safemath{\bimDelta}{\biDelta}
\safemath{\bimLambda}{\biLambda}
\safemath{\bimPhi}{\biPhi}
\safemath{\bimSigma}{\biSigma}
\safemath{\bimOmega}{\biOmega}
\safemath{\bimTheta}{\biTheta}

\safemath{\setA}{\mathcal{A}}
\safemath{\setB}{\mathcal{B}}
\safemath{\setC}{\mathcal{C}}
\safemath{\setD}{\mathcal{D}}
\safemath{\setE}{\mathcal{E}}
\safemath{\setF}{\mathcal{F}}
\safemath{\setG}{\mathcal{G}}
\safemath{\setH}{\mathcal{H}}
\safemath{\setI}{\mathcal{I}}
\safemath{\setJ}{\mathcal{J}}
\safemath{\setK}{\mathcal{K}}
\safemath{\setL}{\mathcal{L}}
\safemath{\setM}{\mathcal{M}}
\safemath{\setN}{\mathcal{N}}
\safemath{\setO}{\mathcal{O}}
\safemath{\setP}{\mathcal{P}}
\safemath{\setQ}{\mathcal{Q}}
\safemath{\setR}{\mathcal{R}}
\safemath{\setS}{\mathcal{S}}
\safemath{\setT}{\mathcal{T}}
\safemath{\setU}{\mathcal{U}}
\safemath{\setV}{\mathcal{V}}
\safemath{\setW}{\mathcal{W}}
\safemath{\setX}{\mathcal{X}}
\safemath{\setY}{\mathcal{Y}}
\safemath{\setZ}{\mathcal{Z}}
\safemath{\emptySet}{\varnothing}

\safemath{\colA}{\mathscr{A}}
\safemath{\colB}{\mathscr{B}}
\safemath{\colC}{\mathscr{C}}
\safemath{\colD}{\mathscr{D}}
\safemath{\colE}{\mathscr{E}}
\safemath{\colF}{\mathscr{F}}
\safemath{\colG}{\mathscr{G}}
\safemath{\colH}{\mathscr{H}}
\safemath{\colI}{\mathscr{I}}
\safemath{\colJ}{\mathscr{J}}
\safemath{\colK}{\mathscr{K}}
\safemath{\colL}{\mathscr{L}}
\safemath{\colM}{\mathscr{M}}
\safemath{\colN}{\mathscr{N}}
\safemath{\colO}{\mathscr{O}}
\safemath{\colP}{\mathscr{P}}
\safemath{\colQ}{\mathscr{Q}}
\safemath{\colR}{\mathscr{R}}
\safemath{\colS}{\mathscr{S}}
\safemath{\colT}{\mathscr{T}}
\safemath{\colU}{\mathscr{U}}
\safemath{\colV}{\mathscr{V}}
\safemath{\colW}{\mathscr{W}}
\safemath{\colX}{\mathscr{X}}
\safemath{\colY}{\mathscr{Y}}
\safemath{\colZ}{\mathscr{Z}}

\safemath{\opA}{\mathbb{A}}
\safemath{\opB}{\mathbb{B}}
\safemath{\opC}{\mathbb{C}}
\safemath{\opD}{\mathbb{D}}
\safemath{\opE}{\mathbb{E}}
\safemath{\opF}{\mathbb{F}}
\safemath{\opG}{\mathbb{G}}
\safemath{\opH}{\mathbb{H}}
\safemath{\opI}{\mathbb{I}}
\safemath{\opJ}{\mathbb{J}}
\safemath{\opK}{\mathbb{K}}
\safemath{\opL}{\mathbb{L}}
\safemath{\opM}{\mathbb{M}}
\safemath{\opN}{\mathbb{N}}
\safemath{\opO}{\mathbb{O}}
\safemath{\opP}{\mathbb{P}}
\safemath{\opQ}{\mathbb{Q}}
\safemath{\opR}{\mathbb{R}}
\safemath{\opS}{\mathbb{S}}
\safemath{\opT}{\mathbb{T}}
\safemath{\opU}{\mathbb{U}}
\safemath{\opV}{\mathbb{V}}
\safemath{\opW}{\mathbb{W}}
\safemath{\opX}{\mathbb{X}}
\safemath{\opY}{\mathbb{Y}}
\safemath{\opZ}{\mathbb{Z}}
\safemath{\opZero}{\mathbb{O}}
\safemath{\identityop}{\opI}

\safemath{\veca}{\bma}
\safemath{\vecb}{\bmb}
\safemath{\vecc}{\bmc}
\safemath{\vecd}{\bmd}
\safemath{\vece}{\bme}
\safemath{\vecf}{\bmf}
\safemath{\vecg}{\bmg}
\safemath{\vech}{\bmh}
\safemath{\veci}{\bmi}
\safemath{\vecj}{\bmj}
\safemath{\veck}{\bmk}
\safemath{\vecl}{\bml}
\safemath{\vecm}{\bmm}
\safemath{\vecn}{\bmn}
\safemath{\veco}{\bmo}
\safemath{\vecp}{\bmp}
\safemath{\vecq}{\bmq}
\safemath{\vecr}{\bmr}
\safemath{\vecs}{\bms}
\safemath{\vect}{\bmt}
\safemath{\vecu}{\bmu}
\safemath{\vecv}{\bmv}
\safemath{\vecw}{\bmw}
\safemath{\vecx}{\bmx}
\safemath{\vecy}{\bmy}
\safemath{\vecz}{\bmz}

\safemath{\veczero}{\bmzero}
\safemath{\vecone}{\bmone}
\safemath{\vecxi}{\bmxi}
\safemath{\veclambda}{\bmlambda}
\safemath{\vecmu}{\bmmu}
\safemath{\vectheta}{\bmtheta}
\safemath{\vecphi}{\bmphi}
\safemath{\vecdelta}{\bmdelta}

\safemath{\matA}{\bA}
\safemath{\matB}{\bB}
\safemath{\matC}{\bC}
\safemath{\matD}{\bD}
\safemath{\matE}{\bE}
\safemath{\matF}{\bF}
\safemath{\matG}{\bG}
\safemath{\matH}{\bH}
\safemath{\matI}{\bI}
\safemath{\matJ}{\bJ}
\safemath{\matK}{\bK}
\safemath{\matL}{\bL}
\safemath{\matM}{\bM}
\safemath{\matN}{\bN}
\safemath{\matO}{\bO}
\safemath{\matP}{\bP}
\safemath{\matQ}{\bQ}
\safemath{\matR}{\bR}
\safemath{\matS}{\bS}
\safemath{\matT}{\bT}
\safemath{\matU}{\bU}
\safemath{\matV}{\bV}
\safemath{\matW}{\bW}
\safemath{\matX}{\bX}
\safemath{\matY}{\bY}
\safemath{\matZ}{\bZ}
\safemath{\matzero}{\bmzero}

\safemath{\matDelta}{\bDelta}
\safemath{\matLambda}{\bLambda}
\safemath{\matPhi}{\bPhi}
\safemath{\matSigma}{\bSigma}
\safemath{\matOmega}{\bOmega}
\safemath{\matTheta}{\bTheta}

\safemath{\matidentity}{\matI}
\safemath{\matone}{\matO}

\safemath{\rnda}{A}
\safemath{\rndb}{B}
\safemath{\rndc}{C}
\safemath{\rndd}{D}
\safemath{\rnde}{E}
\safemath{\rndf}{F}
\safemath{\rndg}{G}
\safemath{\rndh}{H}
\safemath{\rndi}{I}
\safemath{\rndj}{J}
\safemath{\rndk}{K}
\safemath{\rndl}{L}
\safemath{\rndm}{M}
\safemath{\rndn}{N}
\safemath{\rndo}{O}
\safemath{\rndp}{P}
\safemath{\rndq}{Q}
\safemath{\rndr}{R}
\safemath{\rnds}{S}
\safemath{\rndt}{T}
\safemath{\rndu}{U}
\safemath{\rndv}{V}
\safemath{\rndw}{W}
\safemath{\rndx}{X}
\safemath{\rndy}{Y}
\safemath{\rndz}{Z}

\safemath{\rveca}{\bimA}
\safemath{\rvecb}{\bimB}
\safemath{\rvecc}{\bimC}
\safemath{\rvecd}{\bimD}
\safemath{\rvece}{\bimE}
\safemath{\rvecf}{\bimF}
\safemath{\rvecg}{\bimG}
\safemath{\rvech}{\bimH}
\safemath{\rveci}{\bimI}
\safemath{\rvecj}{\bimJ}
\safemath{\rveck}{\bimK}
\safemath{\rvecl}{\bimL}
\safemath{\rvecm}{\bimM}
\safemath{\rvecn}{\bimN}
\safemath{\rveco}{\bomO}
\safemath{\rvecp}{\bimP}
\safemath{\rvecq}{\bimQ}
\safemath{\rvecr}{\bimR}
\safemath{\rvecs}{\bimS}
\safemath{\rvect}{\bimT}
\safemath{\rvecu}{\bimU}
\safemath{\rvecv}{\bimV}
\safemath{\rvecw}{\bimW}
\safemath{\rvecx}{\bimX}
\safemath{\rvecy}{\bimY}
\safemath{\rvecz}{\bimZ}

\safemath{\rvecxi}{\bmxi}
\safemath{\rveclambda}{\bmlambda}
\safemath{\rvecmu}{\bmmu}
\safemath{\rvectheta}{\bmtheta}
\safemath{\rvecphi}{\bmphi}

\safemath{\rmatA}{\bimA}
\safemath{\rmatB}{\bimB}
\safemath{\rmatC}{\bimC}
\safemath{\rmatD}{\bimD}
\safemath{\rmatE}{\bimE}
\safemath{\rmatF}{\bimF}
\safemath{\rmatG}{\bimG}
\safemath{\rmatH}{\bimH}
\safemath{\rmatI}{\bimI}
\safemath{\rmatJ}{\bimJ}
\safemath{\rmatK}{\bimK}
\safemath{\rmatL}{\bimL}
\safemath{\rmatM}{\bimM}
\safemath{\rmatN}{\bimN}
\safemath{\rmatO}{\bimO}
\safemath{\rmatP}{\bimP}
\safemath{\rmatQ}{\bimQ}
\safemath{\rmatR}{\bimR}
\safemath{\rmatS}{\bimS}
\safemath{\rmatT}{\bimT}
\safemath{\rmatU}{\bimU}
\safemath{\rmatV}{\bimV}
\safemath{\rmatW}{\bimW}
\safemath{\rmatX}{\bimX}
\safemath{\rmatY}{\bimY}
\safemath{\rmatZ}{\bimZ}

\safemath{\rmatDelta}{\bimDelta}
\safemath{\rmatLambda}{\bimLambda}
\safemath{\rmatPhi}{\bimPhi}
\safemath{\rmatSigma}{\bimSigma}
\safemath{\rmatOmega}{\bimOmega}
\safemath{\rmatTheta}{\bimTheta}

%% file: standard-macros.tex
\usepackage{amssymb}
\usepackage{amsfonts}
\usepackage{mathrsfs}
\usepackage{xspace}
\usepackage{bm}
\usepackage{fancyref}
\usepackage{textcomp}

\usepackage{multirow}
\usepackage{stmaryrd}

\newenvironment{textbmatrix}{	\setlength{\arraycolsep}{2.5pt}%
								\big[\begin{matrix}}{\end{matrix}\big]%
								\raisebox{0.08ex}{\vphantom{M}}}

\def\be{\begin{equation}}
\def\ee{\end{equation}}
\def\een{\nonumber \end{equation}}
\def\mat{\begin{bmatrix}}
\def\emat{\end{bmatrix}}
\def\btm{\begin{textbmatrix}}
\def\etm{\end{textbmatrix}}

\def\ba#1\ea{\begin{align}#1\end{align}}
\def\bas#1\eas{\begin{align*}#1\end{align*}}
\def\bs#1\es{\begin{split}#1\end{split}} 
\def\bg#1\eg{\begin{gather}#1\end{gather}}
\def\bml#1\eml{\begin{multline}#1\end{multline}}
\def\bi#1\ei{\begin{itemize}#1\end{itemize}}

\newcommand{\lefto}{\mathopen{}\left}

\DeclareMathOperator{\sinc}{sinc}			
\DeclareMathOperator{\Exop}{\opE}			

\newcommand{\Ex}[2]{\ensuremath{\Exop_{#1}\lefto[#2\right]}} 	
\newcommand{\abs}[1]{\lefto\lvert#1\right\rvert}		

\newcommand{\union}{\cup}					

\newcommand{\vecnorm}[1]{\lefto\lVert#1\right\rVert}		
\newcommand{\opnorm}[1]{\lVert#1\rVert}		

\safemath{\dirac}{\delta}					
\safemath{\krond}{\dirac}					

\safemath{\upto}{\uparrow}
\safemath{\downto}{\downarrow}
\safemath{\iu}{j}							
\safemath{\ev}{\lambda}						
\safemath{\hilseqspace}{l^{2}}				
\newcommand{\banachfunspace}[1]{\setL^{#1}}	
\safemath{\hilfunspace}{\banachfunspace{2}}	

\safemath{\SNR}{\textsf{SNR}} 				
\safemath{\PAR}{\textsf{PAR}} 				
\safemath{\No}{N_0}							
\safemath{\Es}{E_s}							
\safemath{\Eb}{E_b}							
\safemath{\EbNo}{\frac{\Eb}{\No}}
\safemath{\EsNo}{\frac{\Es}{\No}}

\DeclareMathOperator{\CHop}{\ensuremath{\opH}} 
\safemath{\tvir}{\rndh_{\CHop}}				
\safemath{\tvtf}{\rndl_{\CHop}}				
\safemath{\spf}{\rnds_{\CHop}}				
\safemath{\bff}{H_{\CHop}}					

\safemath{\ircf}{r_{h}}						
\safemath{\tftvcf}{r_{s}}					
\safemath{\tfcf}{r_{l}}						
\safemath{\bfcf}{r_{H}}						

\safemath{\tcorr}{c_h}						
\safemath{\scf}{c_{s}}						
\safemath{\tfcorr}{c_{l}}					
\safemath{\fcorr}{c_{H}}						

\safemath{\mi}{I}							
\safemath{\capacity}{C}						

\safemath{\normal}{\mathcal{N}}			
\safemath{\jpg}{\mathcal{CN}}			
\safemath{\mchain}{\leftrightarrow}		

\safemath{\dB}{\,\mathrm{dB}}
\safemath{\dBm}{\,\mathrm{dBm}}
\safemath{\Hz}{\,\mathrm{Hz}}
\safemath{\kHz}{\,\mathrm{kHz}}
\safemath{\MHz}{\,\mathrm{MHz}}
\safemath{\GHz}{\,\mathrm{GHz}}
\safemath{\s}{\,\mathrm{s}}
\safemath{\ms}{\,\mathrm{ms}}
\safemath{\mus}{\,\mathrm{\text{\textmu}s}}
\safemath{\ns}{\,\mathrm{ns}}
\safemath{\ps}{\,\mathrm{ps}}
\safemath{\meter}{\,\mathrm{m}}
\safemath{\mm}{\,\mathrm{mm}}
\safemath{\cm}{\,\mathrm{cm}}
\safemath{\m}{\,\mathrm{m}}
\safemath{\W}{\,\mathrm{W}}
\safemath{\mW}{\, \mathrm{mW}}
\safemath{\J}{\,\mathrm{J}}
\safemath{\K}{\,\mathrm{K}}
\safemath{\bit}{\,\mathrm{bit}}
\safemath{\nat}{\,\mathrm{nat}}

\safemath{\define}{\triangleq}			

\safemath{\equivalent}{\sim}
\safemath{\distas}{\sim}					
\safemath{\sdiff}{\Delta}				

\safemath{\reals}{\mathbb{R}}
\safemath{\positivereals}{\reals_{+}}
\safemath{\integers}{\mathbb{Z}}
\safemath{\posint}{\integers_{+}}
\safemath{\naturals}{\mathbb{N}}
\safemath{\posnaturals}{\naturals_{+}}
\safemath{\complexset}{\mathbb{C}}
\safemath{\rationals}{\mathbb{Q}}

\newcommand*{\fancyrefapplabelprefix}{app}		
\newcommand*{\fancyrefthmlabelprefix}{thm}		
\newcommand*{\fancyreflemlabelprefix}{lem}		
\newcommand*{\fancyrefcorlabelprefix}{cor}		
\newcommand*{\fancyrefdeflabelprefix}{def}		
\newcommand*{\fancyrefproplabelprefix}{prop}	
\newcommand*{\fancyrefobslabelprefix}{obs}		
\newcommand*{\fancyrefalglabelprefix}{alg}		
\newcommand*{\fancyrefasmlabelprefix}{asm}	    

\frefformat{vario}{\fancyrefseclabelprefix}{Sec.~#1}
\frefformat{vario}{\fancyrefthmlabelprefix}{Theorem~#1}
\frefformat{vario}{\fancyreflemlabelprefix}{Lemma~#1}
\frefformat{vario}{\fancyrefcorlabelprefix}{Corollary~#1}
\frefformat{vario}{\fancyrefdeflabelprefix}{Definition~#1}
\frefformat{vario}{\fancyrefobslabelprefix}{Observation~#1}
\frefformat{vario}{\fancyrefasmlabelprefix}{Assumption~#1}
\frefformat{vario}{\fancyreffiglabelprefix}{Fig.~#1}
\frefformat{vario}{\fancyrefapplabelprefix}{Appendix~#1} 
\frefformat{vario}{\fancyrefproplabelprefix}{Proposition~#1}
\frefformat{vario}{\fancyrefalglabelprefix}{Algorithm~#1}
\frefformat{vario}{\fancyrefeqlabelprefix}{(#1)}

%% file: defs.tex
\safemath{\dictab}{[\,\dicta\,\,\dictb\,]}

\safemath{\ysig}{\bmy}
\safemath{\ysighat}{\hat{\ysig}}
\safemath{\ysigdim}{M}
\safemath{\xsig}{\bmx}
\safemath{\xsigdim}{N}
\safemath{\nx}{n_x}
\safemath{\zsig}{\bmz}
\safemath{\zsigdim}{\ysigdim}
\safemath{\rsig}{\bmr}
\safemath{\Adict}{\bA}
\safemath{\Adicttilde}{\widetilde{\Adict}}
\safemath{\Adictdim}{\outputdim\times\xsigdim}
\safemath{\avec}{\bma}
\safemath{\avectilde}{\tilde{\avec}}
\safemath{\Bdict}{\bB}
\safemath{\Bdicttilde}{\widetilde{\Bdict}}
\safemath{\Cdict}{\bC}
\safemath{\cvec}{\bmc}
\safemath{\Ddict}{\bD}
\safemath{\Ddictdim}{\ysigdim\times\xsigdim}
\safemath{\dvec}{\bmd}
\safemath{\Ddicttilde}{\widetilde{\bD}}
\safemath{\Bonb}{\bB}
\safemath{\bvec}{\bmb}
\safemath{\Bonbdim}{\ysigdim\times\ysigdim}
\safemath{\noise}{\bmn}
\safemath{\noisedim}{\ysigim}
\safemath{\err}{\bme}
\safemath{\errdim}{\ysigdim}
\safemath{\errset}{\setE}
\safemath{\nerr}{n_e}
\safemath{\delop}{\bP_\errset}
\safemath{\delopc}{\bP_{{\errset}^c}}

\safemath{\cplxi}{\imath}
\safemath{\cplxj}{\jmath}

\safemath{\dict}{\matD}
\safemath{\inputdim}{N}		
\safemath{\outputdim}{M}		
\safemath{\sparsity}{S}	
\safemath{\inputdimA}{{N_a}}	
\safemath{\inputdimB}{{N_b}}	
\safemath{\elemA}{{n_a}}	
\safemath{\elemB}{{n_b}}	
\safemath{\resA}{\matR_a}	
\safemath{\resB}{\matR_b}	
\safemath{\subD}{\matS} 
\safemath{\subA}{\matS_a} 
\safemath{\subB}{\matS_b} 
\safemath{\dicta}{\matA} 	
\safemath{\dictb}{\matB} 	
\safemath{\hollowS}{H}
\safemath{\hollowA}{H_a}
\safemath{\hollowB}{H_b}
\safemath{\cross}{Z}
\safemath{\coh}{\mu_d}			
\safemath{\coha}{\mu_a}			
\safemath{\cohb}{\mu_b}			
\safemath{\mubs}{\nu}	
\safemath{\cohm}{\mu_m} 
\safemath{\dictset}{\setD}	
\safemath{\dictsetp}{\dictset(\coh,\coha,\cohb)}	
\safemath{\dictsetgen}{\dictset_\text{gen}}
\safemath{\dictsetgenp}{\dictsetgen(\coh)}
\safemath{\dictsetonb}{\dictset_\text{onb}}
\safemath{\dictsetonbp}{\dictsetonb(\coh)}

\safemath{\leftside}{U}
\safemath{\rightsideA}{R_a}
\safemath{\rightsideB}{R_b}

\safemath{\indexS}{\setI_S} 

\safemath{\na}{n_a}			
\safemath{\nb}{n_b}			
\safemath{\coeffa}{p_i}	
\safemath{\coeffb}{q_j}	
\safemath{\seta}{\setP}		
\safemath{\setb}{\setQ}     
\safemath{\setw}{\setW}	
\safemath{\setz}{\setZ}	
\safemath{\cola}{\veca}		
\safemath{\colb}{\vecb}		
\safemath{\cold}{\vecd}		
\safemath{\inputvec}{\vecx} 	
\safemath{\error}{\vece}	
\safemath{\noiseout}{\vecz} 	
\safemath{\inputvecel}{x}
\safemath{\inputveca}{\vecx_a}
\safemath{\inputvecb}{\vecx_b}
\safemath{\outputvec}{\vecy}	
\safemath{\lambdamin}{\lambda_{\mathrm{min}}}

\safemath{\elltwo}{\ell_2}
\safemath{\ellone}{\ell_1}
\safemath{\ellzero}{\ell_0}
\safemath{\ellinf}{\ell_\infty}
\safemath{\ellinftilde}{\ell_{\widetilde\infty}}
\safemath{\licard}{Z(\coh,\coha,\cohb)}
\safemath{\xsol}{\hat{x}}
\safemath{\xbord}{x_b}		
\safemath{\xstat}{x_s}		
\safemath{\xstatLone}{\tilde{x}_s}
\safemath{\order}{\mathcal{O}} 
\safemath{\scales}{\Theta} 
\safemath{\ones}{\mathbf{1}} 
\safemath{\zeroes}{\mathbf{0}} 
\safemath{\thlone}{\kappa(\coh,\cohb)} 
\safemath{\constoneA}{\delta} 
\safemath{\constoneB}{\epsilon} 
\safemath{\nlarge}{L}				   
\safemath{\sumlarge}{S_\nlarge}
\safemath{\maxlarger}{P_\nlarge}	   
\safemath{\Pzero}{\textrm{P0}}	
\safemath{\Pone}{\textrm{P1}}
\safemath{\vecfir}{\vecw}			 
\safemath{\vecsec}{\vecz}
\safemath{\elvecfir}{w}              
\safemath{\elvecsec}{z}				 
\safemath{\nlargefir}{n}
\safemath{\normout}{\gamma}
\safemath{\auxfun}{h}
\safemath{\supp}{\textrm{supp}}

\safemath{\indexa}{\ell}
\safemath{\indexb}{r}
\safemath{\indexc}{i}
\safemath{\indexd}{j}

\safemath{\project}{P}